# Resolving the HI in Damped Lyman-α systems that power star-formation

Rongmon Bordoloi[1], John M. O'Meara[2], Keren Sharon[3], Jane R. Rigby[4], Jeff Cooke[5,6], Ahmed Shaban[1], Mateusz Matuszewski[7], Luca Rizzi[2], Greg Doppmann[2], D. Christopher Martin[7], Anna M. Moore[8], Patrick Morrissey[7], & James D. Neill[7]

[1] Department of Physics, North Carolina State University, Raleigh, North Carolina 27695, USA

[2] W.M. Keck Observatory 65-1120 Mamalahoa Highway, Kamuela, Hawaii 96743, USA

[3] Department of Astronomy, University of Michigan, 1085 S. University, Ann Arbor, Michigan 48109, USA

[4] Observational Cosmology Lab, NASA Goddard Space Flight Center, 8800 Greenbelt Rd, Greenbelt, Maryland 20771, USA

[5] Centre for Astrophysics and Supercomputing, Swinburne University of Technology, Hawthorn, Victoria, 3122, Australia

[6] The Australian Research Council Centre of Excellence for All Sky Astrophysics in 3 Dimensions (ASTRO-3D), Hawthorn, Victoria, 3122, Australia

[7] Cahill Center for Astrophysics, Caltech, 1216 East California Boulevard, Pasadena, California 91125, USA

[8] Research School of Astronomy and Astrophysics, Australian National University, Canberra, ACT 2611, Australia

**Reservoirs of dense atomic gas (primarily hydrogen), contain ~90% of the neutral gas at a redshift of 3, and contribute to 2-3% of the total baryons in the Universe[1-4]. These "damped Lyman-α systems" (so called because they absorb Lyman-α photons from within and from background sources) have been studied for decades, but only through**

**absorption lines present in the spectra of background quasars and gamma-ray bursts[5-10]. Such pencil beams do not constrain the physical extent of the systems. Here, we report integral-field spectroscopy of a bright, gravitationally lensed galaxy at a redshift of 2.7 with two foreground damped Lyman-$\alpha$ systems. These systems are $\gtrsim$ 238 kpc$^2$ in extent, with column densities of neutral hydrogen varying by more than an order of magnitude on $\lesssim$ 3 kpc-scales. The mean column densities are ~ $10^{20.46} - 10^{20.84}$ cm$^{-2}$ and the total masses are $\gtrsim 5.5 \times 10^8 - 1.4 \times 10^9\ M_\odot$, showing that they contain the necessary fuel for the next generation of star formation, consistent with relatively massive, low-luminosity primeval galaxies at redshifts > 2.**

## Main

We use the Keck Cosmic Web Imager (*KCWI*)[11] installed on the W. M. Keck Observatory to observe the gravitationally lensed arc SGASJ152745.1+065219. This arc is formed owing to strong gravitational lensing of a galaxy at $z \sim 2.762$, highly magnified and stretched by a massive galaxy cluster at $z \sim 0.392$[12]. Magellan/MagE echelle spectroscopy[13] of the brightest knot of this arc reveals the presence of multiple intervening absorption systems, including a redshift $z = 2.543$ system showing CIV and damped Lyman-$\alpha$ absorption[14]. We perform *KCWI* observations to spatially map out the damped Lyman-$\alpha$ system (DLA) along the arc. The *KCWI* observations reveal the presence of a second DLA at $z \sim 2.05$, at a redshift where the MagE/Magellan observations detect intervening C IV absorption.

Figure 1 shows a spatial map of neutral hydrogen column density ($N_{HI}$) in the $z \sim 2.5$ DLA. We extract moderate signal-to-noise ratio (SNR) spectra for six distinct, spatially resolved positions across the 7 arcsecond spatial extent of SGASJ152745.1+065219 (see Methods). These spectra facilitate a direct analysis of the spatial variation in $N_{HI}$ and the amount of absorption by various heavy elements for the two DLAs along the six lines of sight. Spatially resolved maps of neutral hydrogen and heavy element atomic transitions in high redshift galaxies are highly complementary to lensing tomography of MgII absorption at intermediate redshifts[15,16], quasar lens statistics[17,18] and spatially resolved studies of galactic outflows[19].

Several key features of the $z \sim 2.5$ DLA are immediately visible from Figure 1: (a) strong damped hydrogen absorption is detected at every pointing along the arc, (b) the $N_{HI}$ varies monotonically by an order of magnitude (log $N_{HI} \approx$ 19.9-20.8 cm$^{-2}$) along the east—west extent of the arc, corresponding to $\approx 1$ dex in $N_{HI}$ column density variation over 2-3 kpc in the absorber source-plane (see Methods), (c) the Lyman-$\alpha$ feature trough reaches zero flux (except in aperture F), indicating total, or nearly total coverage of the area illuminated by the background galaxy, (d) the non-zero Lyman-$\alpha$ absorption trough in aperture F with damping wings is indicative of partial DLA coverage, which in itself helps add a limit to the DLA size.

Along with $N_{HI}$ column density variation, we observe large variations in metal absorption line strengths within the $z \sim 2.5$ DLA. Figure 2 shows the variation in absorption strengths of $N_{HI}$, neutral oxygen (OI) and ionized CIV in the source plane of the absorber. The OI absorption strengths along this DLA varies by almost a factor of two, with apertures B and D showing the highest absorption strengths and aperture F showing a non-detection. Different ionization states exhibit significantly different absorption within the same DLA system. The DLA shows large variation in both $N_{HI}$ (<log $N_{HI}$> $\approx 20.46 \pm 0.32$ cm$^{-2}$) and metal absorption strengths and exceeds our conservative 1$\sigma$ errors on the individual HI estimate of 0.2 dex (Extended Data Table 1). The DLA is extended along the declination direction by $\Delta\delta \gtrsim$ 15.9±1.4 kpc and along right ascension by $\Delta\alpha \gtrsim 6.9\pm1.5$ kpc. Each extraction box samples areas of $\approx$ 2–7 kpc$^2$ (depending on where the box lies relative to the regions of strongest magnification) at the redshift of the $z \sim 2.5$ DLA. Such continuous map of spatial variation in an individual DLA is not possible to observe with quasar lines of sight studies as quasars are parsec-scale skewers and multiple quasars appearing sufficiently close together on the sky are extremely rare. The spatial scales probed in this work are similar in size to those probed by very high-resolution simulations of galaxies and the circumgalactic medium[20], providing constraints on the small-scale structure and sizes of individual neutral and ionized CGM gas. These constraints can be used independently to test the subgrid feedback and star-formation prescriptions as well as physical models of gas inflow, outflow, and recycling implemented in simulations[20,21].

As the Lyman-$\alpha$ troughs in most spectra are optically thick (i.e., little to no by-passing flux), we can constrain the minimum size and mass of the DLA by estimating the size of the area covered by the six intervening lines of sight. Such fundamental quantities are largely unconstrained[22,23]. Aperture F is a particularly interesting sightline, as the damped absorption wings confirm it as a DLA, but the absorption trough shows significant by-passing flux ($\gtrsim$ 75% of aperture area covers the background galaxy). This is an example of partial coverage of a DLA in front of an extended background source and validates the technique of using extended background sources to assess DLA sizes via their partial $N_{HI}$ coverage[23]. The maximum extent of the extraction aperture boxes in the six sightlines is $d = 17.4 \pm 0.6$ kpc, corresponding to a minimum projected area of $\gtrsim 238$ kpc$^2$, assuming a circular geometry. We stress that this estimate is a lower limit, as the DLA could extend farther than the area mapped by the background light of SGASJ152745.1+065219, but has the tantalising prospect that we are seeing the edge of at least part of the DLA in Aperture F. The column density $<\log N_{HI}> \approx 20.46 \pm 0.32$ cm$^{-2}$ from an average of the separate estimates provided by each extraction box and a circular uniform slab geometry yields a total neutral hydrogen gas mass of $\gtrsim 5.5^{+0.6}_{-0.3} \times 10^8$ $M_\odot$. Again, this estimate is a strict lower limit for the same reasons as the estimate of the spatial extent and carries the same lensing model uncertainties.

Figure 2 b shows the spatial variation of the second $z \sim 2.055$ DLA, in its source plane. This DLA has higher average HI column density ($<\log N_{HI}> \approx 20.84 \pm 0.02$ cm$^{-2}$) and shows much smaller $N_{HI}$ variation as compared to the former DLA. The variation in metal line absorption strength across the DLA is also more uniform compared to the former, suggesting a larger size with more uniform gas distribution. None of the absorption troughs show signs of partial coverage for this DLA (See Methods). Adopting a circular geometry of the HI across the entire absorber (maximum extent of the apertures $d$ = 18.02 ± 0.56 kpc) yields a projected area ≳ 255 kpc$^2$ at $z$ = 2.0556 and a total neutral hydrogen gas mass ≳$1.4^{+1.5}_{-0.7}$ ×10$^9$ $M_\odot$. We stress that these are strictly lower limits for this DLA.

In the local Universe, extended HI disks are typically seen around galaxies up to a few kpc scales[24], with a small population of disk galaxies showing extended HI disks out to tens of kpc[25]. It is also found that the HI kinematics in local HI disk galaxies[26] are two times smaller than typical DLA kinematics. Our results show that at $z$ > 2, the two DLAs have a spatial extent ≳17 kpc, although the typical halo masses of disk galaxies are much larger at $z$ ~ 0 as compared to $z$ > 2. This suggests that high-$z$ DLAs might have sizes larger than the extended HI disks seen around local galaxies.

We further explore the two DLA systems by extracting a narrow region in wavelength from the *KCWI* cube corresponding to ±3.5 Å around the DLA line centre (See Methods). At these wavelengths, the DLA absorption line centre should absorb all the light from SGASJ152745.1+065219, if the optically thick DLA gas covers the background galaxy light source completely. Yet, as seen in Figure 3 (panels b, c), faint, extended emission is detected in the absorption troughs of both DLAs in a spectrum extracted from the aperture marked in panel a. This emission is particularly strong for the DLA at $z$ ~ 2.05. Some of this emission may stem from the continuum flux by the $z$ ~ 0.4 foreground galaxy to the north of SGASJ152745.1+065219. Two options remain for the remaining flux which is well separated from the foreground galaxy: either the DLA gas only partially covers the background UV light from the continuum of SGASJ152745.1+065219, or it is Lyman-$\alpha$ emission originating from the galaxies associated with the DLAs. We favour the latter interpretation, as a weak emission feature is seen in each extraction box. Moreover, the emission is slightly spatially offset from the brightest background light from SGAS J152745.1+065219 and is seen as small blue clumps in Hubble Space Telescope (*HST)*/ F475W image at these positions (Figure 3). Thus, we interpret the emission as faint Lyman-$\alpha$ emission from star-formation in the two galaxies that host the $z \sim 2.543$ and $z \sim 2.055$ DLAs, seen in absorption and magnified by the gravitational lensing.

Searches for DLA galaxies in emission have yielded few results[27], with the general DLA population shown to be only weakly star forming[28]. Lyman-$\alpha$ emission is also seen in the extracted DLA spectra themselves, offset by $\approx -150\text{-}300$ km/s of the DLA line centre (Methods). One possible explanation for the offset Lyman-$\alpha$ is the association of the Lyman-$\alpha$ emitting gas with inflowing material onto the DLA. Another possibility is that the DLA gas is rotating, with non-uniform star formation throughout. Higher resolution observations with higher signal-to-noise ratio are needed to better constrain the nature of the putative DLA emission. Figure 4 shows location of all pixels with $>3\sigma$ surface brightness levels ($>1.9\times10^{-17}$ erg cm$^{-2}$ s$^{-1}$ arcsec$^{-2}$) in the source plane of the two DLAs. For comparison the six positions at which DLA absorption are measured, are also shown. The absorption-line probes are at very close impact parameter to the DLA hosts, and the hosts have a size typical of a Lyman break galaxy (LBG) at these redshifts[29,30].

# Figures

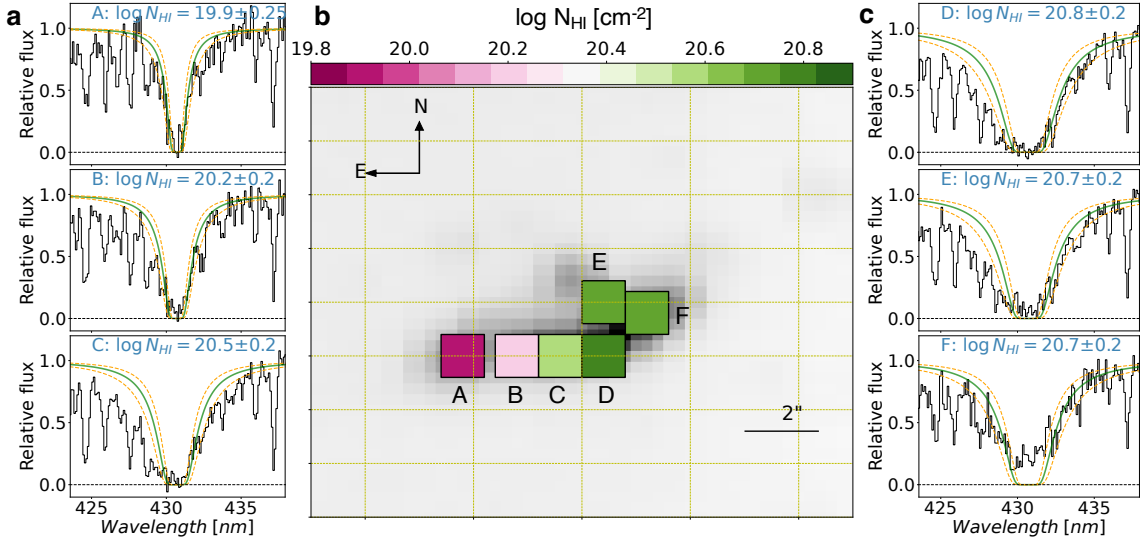

**Figure 1: Spatial variation of neutral hydrogen column density. a**, Extracted $z \sim 2.5$ DLA absorption and the Voigt profile fits used to constrain HI column densities for extraction boxes A-C. The DLA column densities noted in each panel are in units of atoms cm$^{-2}$. Errors quoted are $\pm 1\sigma$ uncertainties in column densities. **b**, Keck/*KCWI* white light image of the background lensed galaxy at $z \sim 2.7$ is shown in grey. The filled squares mark the apertures used in 1D spectral extractions that are used to measure the absorption line properties. The squares are colour coded by their neutral hydrogen column densities in units of atoms cm$^{-2}$. The scalebar corresponds to 2″ in the image plane. The extraction boxes are chosen in the image plane, which correspond to a spatial separation of $\approx$ 2-3 kpc in the source plane of the absorber (see Methods). Over the $\approx$ 2-3 kpc spatial separation, the DLA column densities vary by ~0.9 dex. **c**, Extracted $z \sim 2.5$ DLA absorption and the Voigt profile fits used to constrain HI column densities for extraction boxes D-F.

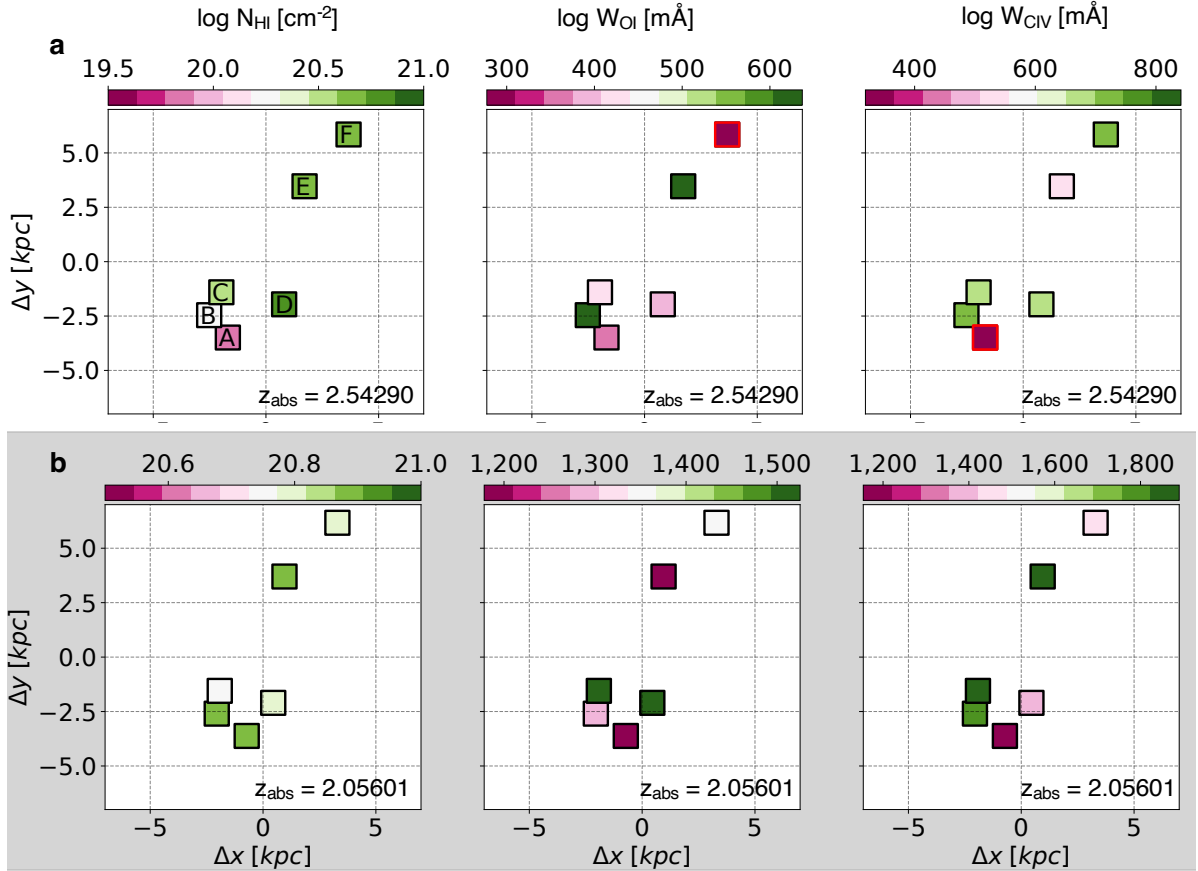

**Figure 2: Maps of HI and metal absorption strengths in the absorber source-plane. a**, Variation of the HI column density, OI absorption strength and CIV absorption strength across the $z \sim 2.5$ DLA in its source plane, respectively. In each case the filled squares are colour coded as a function of their absorption strengths. The squares marked with red outlines are $2\sigma$ upper limits (See Table 1). The squares are marked A-F to identify their location in the image plane (Figure 1). **b**, Same as **a**, but for the $z \sim 2.05$ DLA.

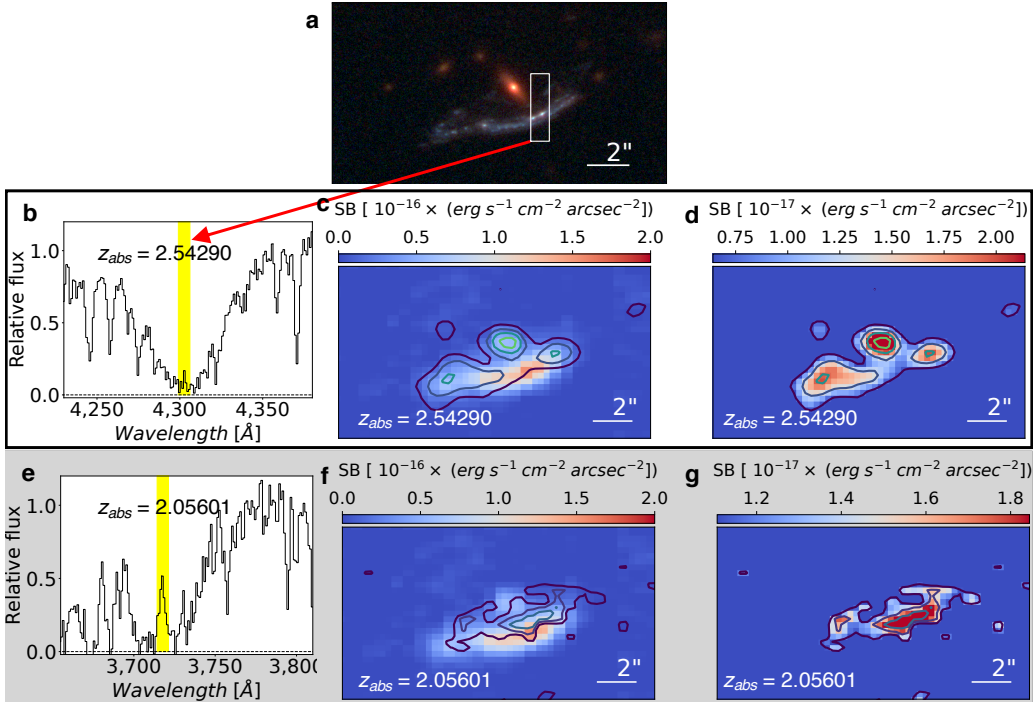

**Figure 3: Lyman-α emission maps of the DLA host galaxies. a**, Colour image of the gravitationally lensed galaxy SGASJ152745.1+065219 rendered from HST/WFC3 F475W (blue), F606W (green), and F110W (red) filters respectively. The rectangle marks the aperture from which extracted spectra are shown in **b** & **e**. **b**, Extracted 1D absorption profile of the foreground ($z = 2.543$) DLA. Yellow patch marks the Lyman-α emission seen in the 1D spectrum. **c,** Continuum emission from the background lensed galaxy ($z = 2.762$) in false colour, with contours marking the ($> 3\sigma$ surface brightness level) of observed Lyman-α emission from the foreground $z = 2.543$ DLA. **d,** Lyman-α emission from the foreground $z = 2.543$ DLA ($>3\sigma$ surface brightness levels), obtained by summing up the emission flux of the Keck/*KCWI* data cube around $1212\ Å \leq \lambda \leq 1219\ Å$ of the rest frame wavelengths of the DLA. **e/f/g**, Same as above but for the foreground $z \sim 2.055$ DLA. The Lyman-α emission from the foreground DLAs are spatially offset from the background galaxy and extend out to at least the size of the background galaxy itself. Note that the surface brightness scales are different in the different panels.

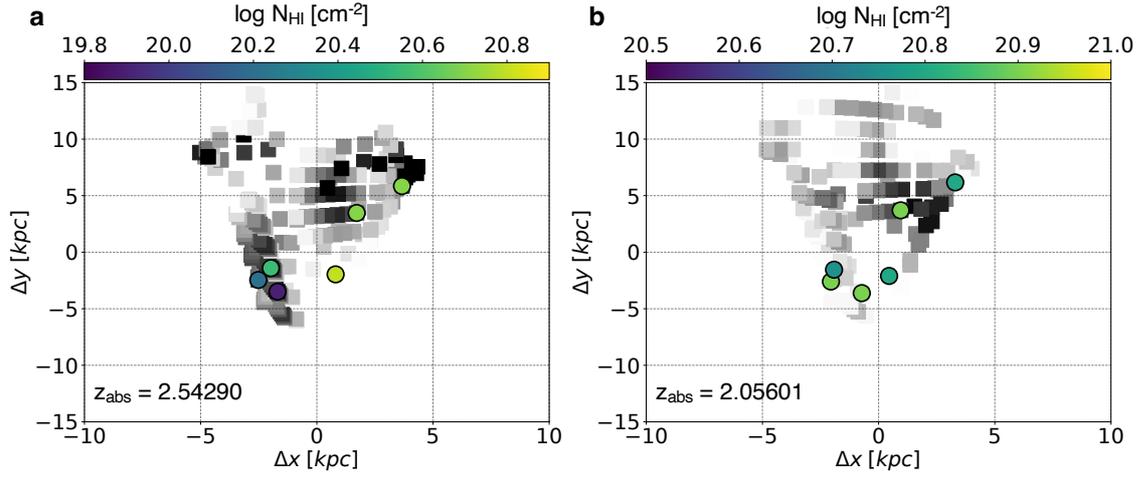

**Figure 4: Lyman-α emission from the background DLAs in their source planes. a**, Source plane positions of pixels that show Lyman-α emission with > 3σ statistical significance in the rest frame of the $z \sim 2.5$ DLA are shown. The filled circles show the DLA column density estimates obtained from absorption line spectroscopy (Figure 1). The background lines of sight intersect the DLA host galaxy at very small (< 1-2 kpc) impact parameter. **b**, Same as the left panel, but for the DLA at $z \sim 2.055$. Lyman-α emission from both the DLA host galaxies have spatial extent similar to typical Lyman-α emitters seen at these redshifts.

**METHODS**

***KCWI* Observations and Data Reduction.**

We observed SGAS J152745.1+065219 with *KCWI* for 2 hours during the night of June 2, 2019. The instrument was configured with the Medium IFU slicer and the BL grating. This combination gives a 16.5″ × 20.4″ field-of-view, a slicer width of 0.69″, and a spectral resolution of $R \approx 2000$. We obtained multiple exposures with exposure times between 600 and 1800 seconds each. Between each exposure, the telescope was shifted by a few arc seconds in the E-W direction to facilitate sky subtraction. The raw data were processed using the publicly available *kderp*[31] package developed by the *KCWI* team, which produces a wavelength- and flux-calibrated data cube for each exposure. Flux calibration was enabled by observing the flux standard Feige 67. We note that, unlike the *HST* data, *KCWI* does not have square pixels on the sky. To best facilitate visual comparison with the *HST* data (as is displayed in Figures 1 and 3), we resampled and optimally combined the *KCWI* data onto square pixels utilizing the *Montage* package[32]. Sky subtraction was performed globally on the cube by median sampling emission-free regions in the data at each wavelength and subtracting that median value from the image.

***HST* Observations and Lens Model**

SGASJ152745.1+065219 was observed by *HST* in Cycle 20 as part of the Sloan Giant Arcs Survey, GO-13003 (PI: Gladders). All observations took place on UT 2013-01-16, with 1211s in WFC3/IR F160W, 1111s in WFC3/IR F110W, and 2400s each in WFC3/UVIS F606W and F475W. The data were reduced using the *AstroDrizzle* package[33], taking care to correct CTE degradation effects, IR blobs, and other artifacts (full description of the data reduction in this program is available in previous papers)[34,35]. Images from the four filters were aligned to the same pixel frame of 0.03″ per pixel (Extended Data Figure 1). The *HST* resolution and broad wavelength coverage allows detection of substructure in the main arc on very small spatial scales. SGAS J152745.1+065219 appears as three partial images of a background galaxy, each marked with a box in Extended Data Figure 1. Due to an alignment of this galaxy and the foreground lens, the central region of the galaxy is highly magnified and lensed into three images. The outskirts of the galaxy are not multiply-imaged and appear only once in the image plane.

The lens model is computed using the public software *Lenstool*[36] and described in detail in a previous publication[34]. The lensing potential in this sightline is dominated by a cluster of galaxies at redshift $z = 0.392$, as measured from spectroscopic redshifts of 14 cluster-member galaxies[37]. Moreover, *KCWI* spectroscopy of the field identifies at least 3 galaxies at $z \sim 0.43$. The lensed galaxy SGAS J152745.1+065219 that is the focus of this study is likely lensed by massive halos on multiple lens planes along the line of sight. Given the proximity of the lens planes, we compute the lens model by projecting the lenses onto the same plane. The cluster lensing potential is constrained by multiple images of five background sources. After solving for the cluster mass distribution, we fix its parameters and treat the cluster as providing external shear when solving for the local lensing perturbation from the elliptical galaxy nearest to SGAS J152745.1+065219. The elliptical galaxy is parameterized as pseudo-isothermal ellipsoidal mass distribution[36], fixed to the position of the observed galaxy, and its ellipticity, position angle, cut radius, core radius, and velocity dispersion are allowed to vary. The bright emission knots and the overall symmetry of SGASJ152745.1+065219 are used as constraints. The best set of parameters is identified in a Markov Chain Monte Carlo process for both the cluster and galaxy halo. The source is reconstructed by ray-tracing the pixels from the *HST* image through the best-fit lens model, as can be seen in Extended Data Figure 1, panel b. One hundred Markov Chain Monte Carlo posterior realizations are used to quantify the $1\sigma$ uncertainty in distances in the absorber planes.

Although the lensing geometry, multiplicity, and distortion of SGASJ152745.1+065219 are a result of lensing from the nearby galaxy, the overall lensing potential in this sight line is dominated by the cluster. Unfortunately, the cluster itself is not well constrained. SGAS J152745.1+065219 is the brightest and most robust lensing constraint in this field, and the other lensed sources that were used to constrain the cluster lensing potential have not been spectroscopically confirmed. This increases the uncertainty on the model outputs[34]. As a result, we place a high uncertainty of a factor of 2 on the lensing magnification of SGASJ152745.1+065219, estimated from a range of models of the foreground cluster. Despite the high magnification uncertainty, the lensing geometry and multiplicity, as well as the source reconstruction, are all well-understood.

**Spectral Extraction**

We developed a custom spectral extraction pipeline to separate variable sized apertures from the *KCWI* cube[38]. There is an astrometric offset between the *KCWI* and *HST* astrometric solutions. We first offset the *KCWI* astrometric solution to match the *HST* observations by offsetting the *KCWI* astrometry by -0.24 arcsecond along the RA and 1.8 arcsecond along the DEC direction, respectively. We then select six spatially separated regions on the *KCWI* white light image (Figure 1) to extract 1D spectra from the *KCWI* data-cube. Our choice of 5×5 spatial pixel box size is made to best balance the signal-to-noise ratio of the extracted spectrum and the desire for small spatial sampling across the SGASJ152745.1+065219 background source. The extracted aperture size exceeds the size of the Full width half maximum (*FWHM*) ≈ 0.8″ seeing conditions during the observations with *KCWI*. For each extraction box, we perform a light weighted optimal extraction by weighting each pixel of the extraction box by the white light image of the data cube. This results in maximum signal-to-noise-ratio for the extracted one-dimensional spectra.

**Absorber modelling & physical characterization of the DLA gas**

Because of the large wavelength extent of the absorption feature, the extracted *KCWI* spectra are of sufficient resolution and signal to noise to perform a Voigt profile analysis of the HI absorption. Using the *linetools*[39] suite of routines, we place a single HI absorption line at $z = 2.543$ and at $z = 2.055$ and vary the HI content until an acceptable model to the data is obtained. Metal absorption lines associated with the two DLAs are single absorption component clouds, justifying the use of a single HI component in the fit. Continuum placement is chosen to match the flux of SGASJ152745.1+065219 on either side of the strong DLA absorption feature. The best fit HI value is obtained by performing *chi squared* minimization using the Levenberg–Marquardt algorithm. The column density uncertainties presented are $1\sigma$ uncertainties on the model fit. In Figure 1, we show the adopted fits on the six extraction boxes for the $z = 2.543$ DLA. Figure 2 and Extended Data Table 1 summarize these measurements.

As the background arc allows the sampling of the same foreground DLAs at different spatial locations, we can study the variation of DLA column density between individual sightlines. There are six unique extracted sightlines piercing the two foreground DLAs, yielding column density differences at 15 distinct physical separations between these sightlines per DLA. Extended Data Figure 2 shows the log $N_{HI}$ variation between 15 separations for each DLAs in their source plane, respectively. It is particularly striking that for the $z = 2.543$ DLA (blue squares), the HI column density varies by an order of magnitude for every 2-3 kpc separation. This dramatic variation suggests the presence of significant small-scale variations within the DLA itself. These values are consistent with our aperture sampling sizes of $\approx 2$–$7$ kpc$^2$ in the source plane of the DLAs. Using broad emission line regions of QSOs it has been shown that proximate DLA clouds near QSOs may be even smaller (<0.32 pc)[40]. However, the current data do not allow the mapping at these spatial scales owing to the trade-off between SNR and seeing. Future observations at either higher SNR, higher spatial resolution (i.e., smaller *KCWI* slicer widths) or both would help explore these smaller scales. This specific example shows the DLA to be a gaseous structure with significant small-scale variation in gas column density and demonstrates the power of this approach of combining IFU observations with gravitational lensing to gain unique insights into such systems.

The spectral resolution of the *KCWI* data is not sufficient to perform a similar Voigt profile analysis on the detected metal absorption lines. Instead, we measure the equivalent width of various ions and adopt the apparent optical depth (AOD) column densities for further analysis. We use a custom absorption line measurement pipeline[41], optimized for CGM absorption line measurements. We perform a local continuum fit around each absorption line by fitting a fourth order Legendre polynomial around ±1000 *km/s* of each absorption line of interest. The $1\sigma$ uncertainty ($\sigma_w$) on each rest frame equivalent width measurement is defined as $\sigma^2_w = (\Sigma\, \sigma^2_\lambda\, \Delta\lambda)/(1+z)$, where $\sigma_\lambda$ is the $1\sigma$ normalized flux uncertainty per pixel, $\Delta\lambda$ is the wavelength difference between pixels and, it is summed over the wavelength range of interest. An absorption line is defined as detected if its rest frame equivalent width is $> 3\sigma_w$. Extended Data Table 1 and Extended Data Table 2 summarize these measurements. Extended Data Figure 3 shows the rest frame equivalent width variations of OI 1302, SiII 1526, SiIV 1393, CII 1334 and CIV 1548 transitions as a function of projected physical separation from the centre of the background arc, respectively. In the $z \sim 2.5$ DLA (blue squares) the low ionization states (OI 1302) and high ionization states (SiIV 1393, CIV 1548) show non-detection in different apertures. OI 1302 transition is not detected along aperture F (Figure 2), which is the sightline with high HI column density, whereas both the SiIV and CIV transitions are not detected along aperture A, the sightline with the lowest HI column density. These variations suggest that the ionization conditions inside the DLA might also be varying.

In all cases the metal absorption line strengths in the $z \sim 2.5$ DLA (blue squares) are ~ 3-4 times weaker than those in the $z \sim 2.05$ DLA (red circles). The $z \sim 2.05$ DLA also exhibits much stronger HI column density than the former DLA. Extending the techniques demonstrated in this paper to a larger sample of spatially resolved DLAs will allow us to rigorously explore the hypothesis of whether low column density DLAs have much more small-scale variations than high column density systems in future works. As OI column densities are saturated in these low-resolution observations, we are not able to estimate accurate OI column densities and therefore cannot measure accurate DLA metallicities. Future works with higher SNR and resolution observations will enable the creation of metallicity maps at tens of parsec scale spatial resolution for this observation.

**Spatially extended Lyman-*α* emission**

The extracted KCWI spectra at different spatial positions along the lensed arc show significant variation in DLA column density (Figure 2). However, when a 1D spectrum is extracted centred on the aperture shown in Figure 3, panel a, a faint emission spike is seen at the core of the DLA absorption for both the DLAs (see Figure 3, panels b, e). The emission is particularly strong for the $z \sim 2.05$ DLA, and we detect strong emission spikes for all the apertures used in this study (Extended Data Figure 4). We explore if this emission is seen on other parts of the sky by summing the flux around ±3.5 Å of Lyman-*α* line centre at $z \sim 2.5$, and $z \sim 2.05$, respectively. Figure 3 (panels d & g) show the statistically significant (>3$\sigma$ surface brightness) Lyman-*α* emission maps for both the DLAs, respectively. Note that the surface brightness colour-bars show different scales for the two DLAs. Panels c & f, show the continuum image for the background lensed galaxy. The DLA Lyman-*α* maps are overlaid in these panels as contours. The contours mark the $3\sigma$ to $15\sigma$ statistically significant surface brightness levels of Lyman-$\alpha$ emission of both the DLAs, respectively. Some emission might come from the foreground cluster galaxy at $z \sim 0.4$, but the extended emission features on either side are not coming from this galaxy. Some of the diffuse emission features may be associated with the blue diffuse emission in the colour composite *HST* image (Figure 3). The corresponding continuum emission from the background lensed galaxy SGAS J152745.1+065219 at $z \sim 2.7$ is shown in panels c and f. It is clearly seen that the Lyman-$\alpha$ emission associated with the foreground DLA (contours) and the background lensed galaxy are not co-spatial. As we do not currently detect the continuum associated with this DLA Lyman-$\alpha$ emission, we constrain the minimum Lyman-$\alpha$ rest frame emission equivalent width to be $> 1.1$ Å for the $z \sim 2.5$ DLA, and $> 3.5$ Å for the $z \sim 2.05$ DLA, respectively. These emission fluxes correspond to limits on host galaxy star-formation rates $> 0.35\ M_\odot\ yr^{-1}$ for the $z \sim 2.5$ DLA, and $> 1\ M_\odot\ yr^{-1}$ for the $z \sim 2.05$ DLA, respectively.

**Methods References**

**Data Availability**

Data that support the findings of this study are publicly available at the Keck Observatory Archive (KOA), https://www2.keck.hawaii.edu/koa/public/koa.php, under project codes N083, K338, and the Barbara A. Mikulski Archive for Space Telescope under project code GO-13003. Fully reduced data are available from the corresponding author upon request.

**Code Availability**

All codes used in this work are publicly available. The HI column density measurements are performed using the *linetools* package[39]. Reduction and analysis of the *KCWI* data cubes were done using the *kcwitools* package[38]. The lensing raytracing and absorption line measurements are done using the *rbcodes* package[41]. *HST* image analysis and lens modelling was performed with *AstroDizzle*[33] software and the *Lenstool*[36], respectively.

**Acknowledgements**

This work was supported by a NASA Keck PI Data Award, administered by the NASA Exoplanet Science Institute. Data presented herein were obtained at the W. M. Keck Observatory from telescope time allocated to the National Aeronautics and Space Administration through the agency's scientific partnership with the California Institute of Technology and the University of California. The Observatory was made possible by the generous financial support of the W. M. Keck Foundation. This research was conducted, in part, by the Australian Research Council Centre of Excellence for All Sky Astrophysics in 3 Dimensions (ASTRO 3D), through project number CE170100013. The authors wish to recognize and acknowledge the very significant cultural role and reverence that the summit of Mauna Kea has always had within the indigenous Hawaiian community. We are most fortunate to have the opportunity to conduct observations from this mountain. This research made use of Montage. It is funded by the National Science Foundation under Grant Number ACI-1440620, and was previously funded by the National Aeronautics and Space Administration's Earth Science Technology Office, Computation Technologies Project, under Cooperative Agreement Number NCC5-626 between NASA and the California Institute of Technology.

**Author Contributions**

R.B. and J.M.O developed the idea for the project, wrote the NASA/Keck telescope proposal and designed and performed the observations. R.B. developed the analysis tools performed the analysis, devised original ways to interpret the results and authored majority of the text. J.M.O reduced the KCWI data. A.S. performed the metal absorption line measurements. K.S. performed the lens model and provided Extended Data Figure 1. J.R.R. provided the ancillary data from MagE and metal absorber information from MagE spectra. J.C., J.M.O. & R.B. provided steps to correct astrometric offsets and J.C. confirmed the redshift of the second DLA, contributed to the interpretations. M.M., L.R., G.D., D.C.M, A.M.M., P.M., J.D.N. developed the KCWI data reduction pipeline and built and delivered the instrument when initial commissioning data provided the data needed to verify the target as an object of interest. All authors, including J.M.O, J.R.R, J.C., contributed to the overall interpretation of the results and various aspects of the analysis and writing.

**Competing interests**

The authors declare that they have no competing financial interests.

**Author Information**

**Department of Physics, North Carolina State University, Raleigh, NC 27695, USA**

Rongmon Bordoloi & Ahmed Shaban

**W.M. Keck Observatory 65-1120 Mamalahoa Highway, Kamuela, Hawaii 96743, USA**

John M. O'Meara, Luca Rizzi & Greg Doppmann

**Department of Astronomy, University of Michigan, 1085 S. University, Ann Arbor, Michigan 48109, USA**

Keren Sharon

**Observational Cosmology Lab, NASA Goddard Space Flight Center, 8800 Greenbelt Rd, Greenbelt, MD 20771, USA**

Jane R. Rigby

**Centre for Astrophysics and Supercomputing, Swinburne University of Technology, Hawthorn, Victoria, 3122, Australia**

**The Australian Research Council Centre of Excellence for All Sky Astrophysics in 3 Dimensions (ASTRO-3D), Hawthorn, Victoria, 3122, Australia**

Jeff Cooke

**Cahill Center for Astrophysics, Caltech, 1216 East California Boulevard, Pasadena, California 91125, USA**

Mateusz Matuszewski, D. Christopher Martin, Patrick Morrissey & James D. Neill

**Research School of Astronomy and Astrophysics, Australian National University, Canberra, ACT 2611, Australia**

Anna M. Moore

**Extended Data Figures**

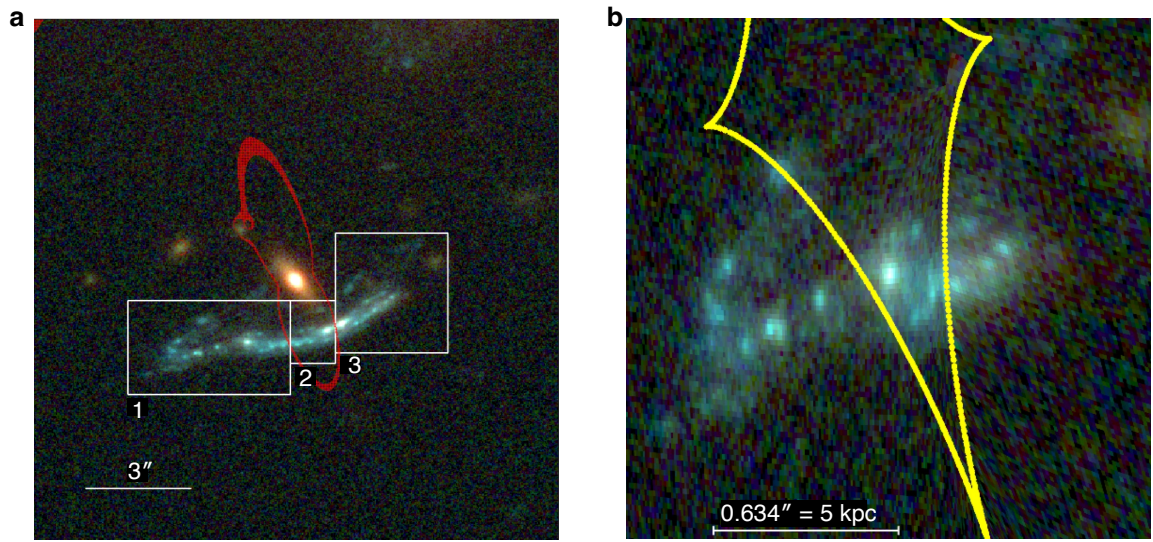

**Extended Data Figure 1: Source plane reconstruction of SGAS J152745.1+065219. a,** WFC3/*HST* images of SGAS J152745.1+065219 in the WFC3-IR F160W, WFC3-UVIS F606W, and WFC3-UVIS F475W filters[34]. North is up and East is to the left. The gravitational lensing critical curve is shown in red, representing areas in the image plane with extreme magnification. Three boxes mark the locations of the three partially lensed images. The lensing potential is caused by the $z = 0.43$ elliptical galaxy at the centre of this field, boosted by a cluster of galaxies at $z = 0.39$, located within 1 arcminute in projection North-East of this galaxy. **b,** The reconstructed source plane image of the galaxy. The source-plane caustic is marked in yellow, representing regions with extreme magnification, and defining the multiplicity of the strongly lensed source. The region interior to the cusp is lensed into three images, whereas areas outside the cusp are magnified, but not multiply imaged.

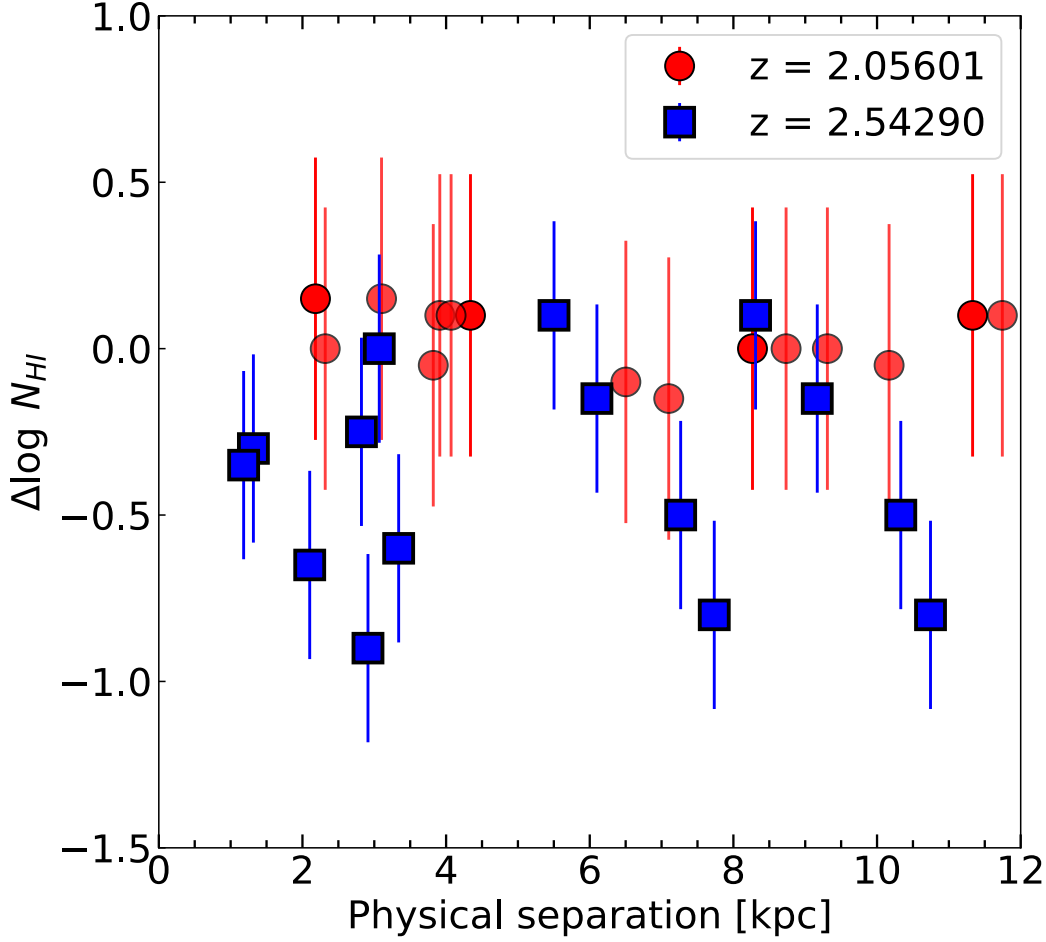

**Extended Data Figure 2: Pairwise HI column density variation vs physical separation.** Variation in HI column density between six individual pointing to the background arc as a function of physical separation between them. Error bars correspond to the ±1σ uncertainty in column density ratios. The $z \sim 2.5$ DLA (blue squares) shows an order of magnitude variation in column density in 2-3 kpc separations. This suggests significant small-scale variation inside the DLA, at 2-3 kpc physical scales. In contrast, the $z \sim 2.05$ DLA (red circles) shows very little variation in column densities across different sightlines. The red circles are offset in x-direction by 1 kpc for clarity of presentation.

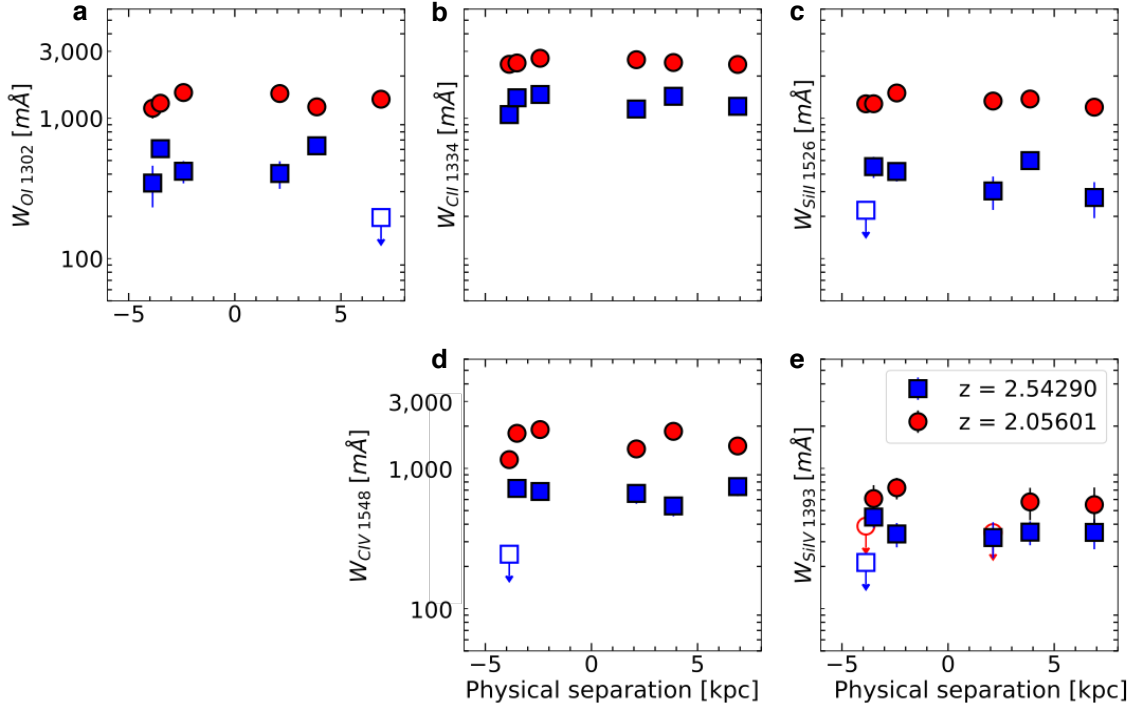

**Extended Data Figure 3: Variation in metal absorption line strengths of the two DLA systems. a-e,** Metal absorption line strength variations of different ions across the arc for the $z \sim 2.5$ DLA (blue squares) and the $z \sim 2.05$ DLA (red circles), respectively. The physical separations are in the source-plane of the absorbers and centred on the centre of the background arc. The filled symbols are detections, and the open symbols are $2\sigma$ limits of non-detections. Error bars correspond to the $\pm 1\sigma$ uncertainty measurement of absorption strengths. In all cases, the $z \sim 2.5$ DLA exhibit much weaker metal absorption lines as compared to the $z \sim 2.05$ DLA.

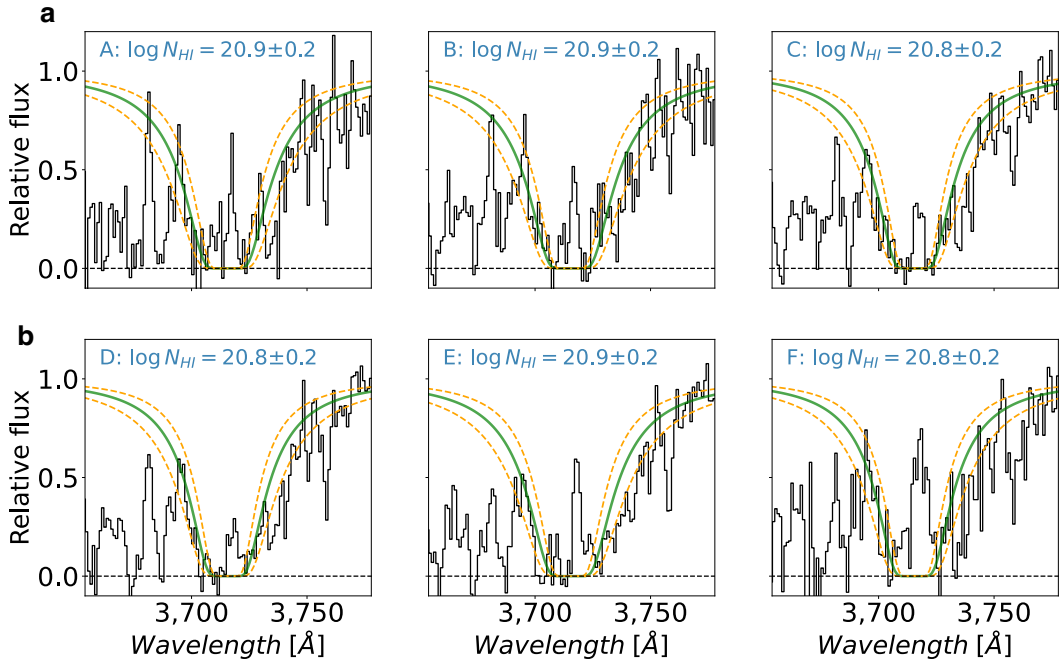

**Extended Data Figure 4: Spatial variation of neutral hydrogen column density of the z ~ 2.05 DLA. a**, The extracted 1D Lyman-$\alpha$ absorption profiles are shown for apertures A-C. The best fit Voigt profiles with ±1$\sigma$ error bounds are shown as solid green and dashed orange lines, respectively. The DLA column densities are marked in each panel and are in units of atoms cm$^{-2}$. In each absorption profile, the DLA absorption trough reaches zero flux, indicating that the aperture is fully covering the DLA gas cloud. The emission spike in the middle of the absorption trough corresponds to Lyman-$\alpha$ emission leaking out of the corresponding DLA host galaxy. **b**, Same as **a**, but for apertures D-F.

**Extended Data Tables**

**Extended Data Table 1: Absorption line measurements for the $z = 2.54290$ DLA.**

| Aperture[a] | log $N_{HI}$[b] | $W_{OI1302}$[c] | $W_{CII1334}$[c] | $W_{SiII1526}$[c] | $W_{CIV1548}$[c] | $W_{SiIV1393}$[c] |
|---|---|---|---|---|---|---|
| A | 19.90±0.2 | 345±114 | 1062±120 | <221 | <244 | <214 |
| B | 20.20±0.2 | 607±69 | 1400±72 | 452±77 | 720±89 | 450±73 |
| C | 20.55±0.2 | 419±76 | 1478±65 | 418±63 | 684±79 | 341±67 |
| D | 20.80±0.2 | 404±90 | 1164±93 | 303±81 | 663±104 | 321±92 |
| E | 20.70±0.2 | 637±74 | 1435±73 | 499±71 | 540±85 | 351±68 |
| F | 20.70±0.2 | <196 | 1217±88 | 272±78 | 741±104 | 350±85 |

Absorption line measurements of the DLA at $z \sim 2.5$. [a] Aperture name as denoted in Figure 1. [b] Measured HI column densities from Voigt profile fitting. Errors are ±1σ uncertainties on column densities. [c] Rest frame equivalent widths of different metal absorption lines in units of mÅ. Errors are ±1σ uncertainties on equivalent widths. The < signs denote 2σ non-detection limits.

**Extended Data Table 2**: **Absorption line measurements for the *z* = 2.05601 DLA.**

| Aperture[a] | log $N_{HI}$[b] | $W_{OI1302}$[c] | $W_{CII1334}$[c] | $W_{SiII1526}$[c] | $W_{CIV1548}$[c] | $W_{SiIV1393}$[c] |
|---|---|---|---|---|---|---|
| A | 20.90±0.3 | 1178±181 | 2425±257 | 1268±123 | 1154±131 | <387 |
| B | 20.90±0.3 | 1280±114 | 2478±168 | 1269±77 | 1778±97 | 609±150 |
| C | 20.75±0.3 | 1525±101 | 2677±151 | 1515±82 | 1890±91 | 728±127 |
| D | 20.80±0.3 | 1501±157 | 2615±214 | 1327±100 | 1376±127 | <349 |
| E | 20.90±0.3 | 1206±96 | 2496±166 | 1372±85 | 1836±99 | 577±150 |
| F | 20.80±0.3 | 1367±153 | 2418±223 | 1200±110 | 1444±124 | 551±180 |

Absorption line measurements of the DLA at *z* ~ 2.05601. [a] Aperture name as denoted in Figure 1. [b] Measured HI column densities from Voigt profile fitting. Errors are ±1σ uncertainties on column densities. [c] Rest frame equivalent widths of different metal absorption lines in units of mÅ. Errors are ±1σ uncertainties on equivalent widths. The < signs denote 2σ non-detection limits.